\documentclass{sig-alternate}
\usepackage{times,amsmath,epsfig}
\usepackage{times,amsmath,epsfig}
\usepackage{amssymb}
\usepackage{amsfonts}
\usepackage{graphicx}
\usepackage{algorithm}
\usepackage{subfigure}
\usepackage{url}
\usepackage[noend]{algorithmic}
\def\mund#1{\smallskip\noindent{\bf #1:}}
\def\mnvp{\vspace*{-0.1in}}
\def\mvp{}

\def\Ignore#1{}

\newenvironment{definition}[1][Definition]{\begin{trivlist}
\item[\hskip \labelsep {\bfseries #1}]}{\end{trivlist}}

\def\SuppInfo {{\sl specificity}}
\def\ConfAFD {{\sl confidence}}
\def\smartint {{\bf {\sc SmartInt}}}

\begin{document}


%
\crdata{}

\newfont{\sttlfnt}{phvb8t at 14pt}
\title{S{\sttlfnt MART}I{\sttlfnt NT}: Using Mined Attribute Dependencies to Integrate
Fragmented Web Databases }

%
%
%
%
%

\numberofauthors{5} 
%

\author{
Ravi Gummadi  \hspace*{0.1in} Anupam Khulbe \hspace*{0.1in}  Aravind
Kalavagattu \hspace*{0.1in}  Sanil Salvi \hspace*{0.1in}  Subbarao
Kambhampati\\
Arizona State University, Tempe AZ,  85287 USA\\
\{gummadi,khulbe,akalavagattu,sdsalvi,rao\}@asu.edu
}

\maketitle
\def\addauflag{1} 

\begin{abstract}
Many web databases can be seen as providing partial and overlapping
information about entities in the world. To answer queries effectively,
we need to integrate the information about the individual entities
that are fragmented over multiple sources. At first blush this is just
the inverse of traditional database normalization problem - rather
than go from a universal relation to normalized tables, we want to
reconstruct the universal relation given the tables (sources).
The standard way of  reconstructing the
entities will involve joining the tables.
Unfortunately, because of the autonomous and decentralized way in
which the sources are populated, they often do not have Primary Key -
Foreign Key relations.
While tables may share attributes, naive joins over these
shared attributes can result in reconstruction of many spurious
entities thus seriously compromising precision. Our system, \smartint\ is aimed at addressing  the problem of data
integration in such scenarios. Given a query,
our system uses the Approximate Functional Dependencies (AFDs) to piece
together a tree of relevant tables to answer it.
The result tuples produced by our system are able to strike a favorable balance between precision and
recall.
\end{abstract}
%
%
\section{Introduction}\label{sec:intro}

With the advent of web, data available online is rapidly increasing,
and an increasing portion of that data corresponds to large number of
web databases populated by web users. Web databases can be viewed as
providing partial but overlapping information about entities in the
world. Conceptually, each entity can be seen as being fully described
by a universal relation comprising of all its attributes. Individual
sources can be seen as exporting parts of this universal relation.
This picture looks very similar to the traditional database set-up.
The database administrator (who ensures lossless normalization) is
replaced by \emph{independent data providers}, and specialized users (who are aware of database
querying language) are replaced by \emph{lay users}. These changes have
two important implications:
\begin{itemize}
\item \textbf{Ad hoc Normalization by providers}:
Primary key-Foreign key (PK-FK) relationships that are crucial for reconstructing the universal
relation are often missing from the tables. This is in part because
partial information about the entities are independently entered by data providers into
different tables, and synthetic keys (such as vehicle ids, model ids,
employee ids) may not be uniformly preserved across sources. (In some cases, such as
public data sources about people,
the tables may even be explicitly forced to avoid keeping such key information.)

\item \textbf{Imprecise queries by lay users}:
Most users accessing these tables are lay users and are often not
aware of all the attributes of the universal relation. Thus their
queries may be ``imprecise'' \cite{AIMQ} in that they may miss requesting some of
the relevant attributes about the entities under consideration.

\Ignore{
\item \textbf{Imprecise (Completion) queries by lay users}: Most users accessing
  these tables are lay users and are often not aware of all the
  attributes of the universal relation. Thus they tend to pose
imprecise queries (i.e., ``{\tt Select *}'' queries that do not
specify all the attributes of interest.}


\end{itemize}

\begin{figure}
\center
\includegraphics[width=3in]{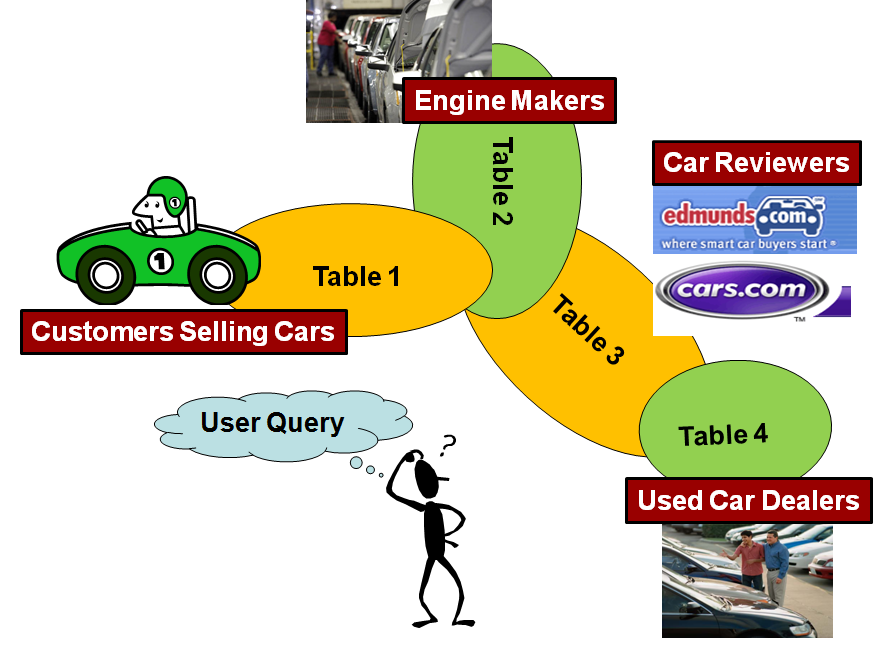}
\caption{Overlapping tables in the database}
\mvp
\label{fig:overlap}
\vspace*{-0.2in}
\end{figure}

%


Thus a core part of the source integration on the web can be cast as
the problem of reconstructing the universal relation in the absence of
primary key-foreign key relations, and in the presence of lay users.
Our main aim in this paper is to provide a fully automated solution to
this problem.  One reason this problem has not received much attention
in the past is that it is often buried under the more immediate
problem of attribute name heterogeneity:  In addition to the loss of PK-FK
information, different tables tend to rename their columns.\footnote{in other
words, web data sources can be seen as resulting from an {\em
ad hoc} normalization followed by the attribute name change} While
many reasonable schema mapping solutions have been developed to handle
the schema heterogeneity problem (c.f.
\cite{melnik:simiflood,halevy:mlinteg,li:semint,larson:theory}), we are not aware of any effective solutions
for the reconstruction problem. In this paper (as well as in our
implemented system) we will simply assume that the attribute name change
problem can be handled by adapting one of  the existing methods. This
allows us to focus on the central problem of reconstruction of universal relation in the
absence of primary key-foreign key relationships.

\subsection{Motivating Scenario}
As a motivating scenario, let us consider a set of tables (with
different schema) populated in a Vehicle domain (Figure \ref{fig:overlap}).
The universal schema
of entity `Vehicle' can be described as follows:
\emph{ Vehicle (VIN, vehicle-type, location, year, door-count, model, make, review, airbags, brakes, year, condition, price, color, engine, cylinders, capacity, power, dealer, dealer-address)}

 Let us assume that the database has the following tables: {\it Table \ref{table:schema1} with Schema S1} - populated by normal web users who sell and buy cars, {\it Table \ref{table:schema2} with Schema S2} - populated by crawling reviews of different vehicles from websites, {\it Table \ref{table:schema3} with Schema S3} - populated by engine manufacturers/vendors with specific details about vehicle engines and {\it Table \ref{table:schema4} with Schema S4}. The following shows the schema for these tables and the corresponding schema mappings among them: \emph{S1 -
(make, model\_name, year, condition, color, mileage, price, location, phone), S2 - (model, year, vehicle-type, body-style, door-count, airbags, brakes, review, dealer), S3 - (engine, mdl, cylinders, capacity, power) and S4 - (dealer, dealer-address, car-models)}

The following attribute mappings are present among the
schema:
{\it (S1: model\_name = S2: model = S3: mdl, S2: dealer = S4: dealer)}
The italicized attribute \emph{MID (Model ID)} refers to a \textit{synthetic primary key} which would have been
present if the users shared understanding
about the entity which they are populating. If it is present,
entity completion becomes trivial because you can simply use that attribute to join the tables. There  can
be a variety of reasons why that attribute is not available:
(1) In autonomous databases, users populating the data are not aware of all the attributes and may end up missing the `key' information.
(2) Since each table is autonomously populated, though each table has a key, it might not be a shared attribute.
(3) Because of the decentralized way the sources are populated, it
is hard for the sources to agree on ``synthetic keys''
(that sometimes have to be generated during traditional
normalization).
(4) The primary key may be   intentionally masked, since it describes
sensitive information about the entity (e.g. social security number).
\vspace*{-0.2in}

\begin{table}[ht]
\caption{\bf Schema 1 - Cars($S_1$)}
\mvp
\centering 
\begin{small}\begin{tabular}{|c |c |c |c |c|} 
\hline 
\emph{MID} & Make& Model\_name&Price&Other Attrbs \\
\hline
\emph{HACC96}&Honda&Accord&19000& \ldots \\
\emph{HACV08}&Honda&Civic&12000& \ldots \\
\emph{TYCRY08}&Toyota&Camry&14500& \ldots \\
\emph{TYCRA09}&Toyota&Corolla&14500& \ldots \\
\hline
\end{tabular}\end{small}
\label{table:schema1} 
\vspace*{-0.3in}
\end{table}

\begin{table}[hbt]
\caption{\bf Schema 2 - Reviews ($S_2$)}
\mvp
\centering 
\begin{small}\begin{tabular}{|c |c |c |c|c|c|} 
\hline 
Model & Review & Vehicle-type & Dealer & Other Attrb\\
\hline
Corolla & Excellent & Midsize & Frank &...\\
Accord & Good & Fullsize & Frank &...\\
Highlander & Average & SUV & John &...\\
Camry & Excellent & Fullsize & Steven & ...\\
Civic & Very Good & Midsize & Frank &...\\
\hline
\end{tabular}\end{small}
\label{table:schema2} 
\vspace*{-0.3in}
\end{table}

\begin{table}[ht]
\caption{\bf Schema 3 - Engine ($S_3$)}
\mvp
\centering 
\begin{small}\begin{tabular}{|c |c |c |c |c|c|c|} 
\hline 
\emph{MID} & Mdl & Engine & Cylinders & Other Attrb\\
\hline
\emph{HACC96} & Accord & K24A4 & 6 & ...\\
\emph{TYCRA08} & Corolla & F23A1 & 4 & ...\\
\emph{TYCRA09} & Corolla & 155 hp & 4 & ...\\
\emph{TYCRY09} & Camry & 2AZ-FE I4 & 6 & ...\\
\emph{HACV08} & Civic & F23A1 & 4 & ...\\
\emph{HACV07} & Civic & J27B1 & 4 & ...\\
\hline
\end{tabular}\end{small}
\label{table:schema3} 
\vspace*{-0.1in}
\end{table}

\begin{table}[ht]
\caption{\bf Schema 4 - Dealer Info ($S_4$)}
\mvp
\centering 
\begin{small}\begin{tabular}{|c |c |c |} 
\hline 
Dealer & Address & Other Attrb\\
\hline
Frank & 1011 E Lemon St, Scottsdale, AZ & ...\\
Steven & 601 Apache Blvd, Glendale, AZ & ...\\
John & 900 10th Street, Tucson, AZ & ...\\
\hline
\end{tabular}\end{small}
\label{table:schema4} 
\vspace*{-0.2in}
\end{table}


\begin{table}[ht]
\caption{\bf Results of Query Q just from Table $T_1$}
\mvp
\centering 
\begin{small}\begin{tabular}
{|c |c |c |} 
\hline
Make&Model&Price\\
\hline
Honda&Civic&12000\\
Toyota&Camry&14500\\
Toyota&Corolla&14500\\
\hline
\end{tabular}\end{small}
\label{table:singletable} 
\vspace*{-0.2in}
\end{table}

\begin{table}[ht]
\caption{\bf Results of Query Q using direct-join ($T1 \bowtie T3$)}
\mvp
\centering 
\begin{small}
\begin{tabular}{|c |c |c | c| c| c|} 
\hline
Make&Model&Price&Cylinder&Engine&Other attrbs\\
\hline
Honda&Civic&12000&4&F23A1 &...\\
Honda&Civic&12000&4&J27B1 &...\\
Toyota&Corolla&14500&4&F23A1&...\\
Toyota&Corolla&14500&4&155 hp&...\\
\hline
\end{tabular}\end{small}
\label{table:resultsjoin} 
\vspace*{-0.2in}
\end{table}


\begin{table}[ht]
\caption{\bf Results of Query Q using attribute dependencies}
\mvp
\centering 
\begin{small}\begin{tabular}{|c |c |c | c| c| c| c|} 
\hline
Make&Model&Price & Cylinders & Review & Dealer & Address\\
\hline
Honda & Civic & 12000 & 4 & Very Good & Frank & 1011 E St\\
Toyota & Corolla & 14500 & 4 & Excellent & Frank & 1011 E St\\
\hline
\end{tabular}\end{small}
\label{table:resultsafd} 
\vspace*{-0.2in}
\end{table}

\Ignore{Suppose the user is interested in the following query:
\textbf{Give me `Make', `Model' of all vehicles whose price is less
than \$15000 and which have a 4-cylinder engine.}\footnote{We use this
example throughout the paper to illustrate the  working of different modules of the system} The above
query would translate to the following SQL notation.}

Consider the following representative queries (that \smartint\ is aimed at handling):

\noindent{\bf Q1}: \texttt{SELECT make, model \\ WHERE price $<$ \$15000 AND cylinders = `4'}.

\smallskip
\noindent{\bf Q2}: \texttt{SELECT make, vehicle-type \\ WHERE price $<$ \$15000 AND cylinders = `4'}.

\noindent{\bf Q3}: \texttt{SELECT  *   \\ WHERE price $<$ \$15000 AND cylinders = `4'}.

\medskip

The first thing to note is that all these queries are {\em partial} in
that they do not specify the exact tables over which the query is to
be run.  Further more, note that both the query constraints and
projected attributes can be are {\em distributed over multiple
  tables}. In Q1, the constraint on price can only be evaluated over
the first table, while the constraint on the number of cylinders can
only be evaluated on table 3. In Q2, the projected attributes are also
distributed across different tables. Finally, Q3 is an imprecise (entity
completion)  query, where the user essentially wants all the
information--spread across different tables--on the entities that
satisfy the constrains.
Before we introduce our approach, let us examine the limitations of two obvious
approaches to answer these types of  queries in our scenario:

\smallskip
 \noindent \textbf{Answering from a single table}: The first approach is
to answer the query from the single table which conforms to the most number
of constraints mentioned in the query and provides maximum number of
attributes. In the given query since `make', `model' and `price' map
onto Table \ref{table:schema1}, we can directly query that table by
ignoring the  constraint on the `cylinders'. The resulting tuples are
shown in Table \ref{table:singletable}. The second tuple related to
`Camry' has 6 cylinders and is shown as an answer. Hence ignoring
constraints would lead to erroneous tuples in the final result set
which do not conform to the query constraints.

\smallskip

\noindent \textbf{Direct  (Naive) Join}: The second and a seemingly more reasonable
approach is joining the tables using whatever shared attribute(s) are
available. The result of doing a direct join based on the shared
attribute(`model') is shown in Table \ref{table:resultsjoin}. If we
look at the results, we can see that even though there is only one
`Civic' in Table \ref{table:schema1}, we have two Civics in the final
results. The same happens for `Corolla' as well. The absence of
Primary Key - Foreign Key relationship between these two tables has
lead to spurious results.

\Ignore{
Apart from the limitations discussed above, these approaches also fail
to get other relevant attributes which describe the entity. In such a scenario, providing the complete information about the entity to users requires:
\begin{itemize}
\item Linking attributes and propagating constraints spanning across multiple tables, and retrieving precise results.
\item Increasing the completeness of the individual results by
retrieving additional relevant attributes and their associated values
from other overlapping tables not specified in the query(thereby
reconstructing the universal relation from different local schema).
\end{itemize}
%
Addressing these two needs poses serious challenges. In the absence of
information on how the tables overlap, it is not possible to link the
attributes across tables. We can find the mappings between attributes using algorithms like Similarity Flooding \cite{melnik:simiflood}. However, these alone would not
be enough. Since attribute mappings between tables still leaves the
problem of absence of key information open, the usual process of
getting results through direct join would result in very low
precision. Moreover discovering additional attributes related to those
mentioned in the query requires the  knowledge of attribute dependencies, which are not apparent.
}

\subsection{\textbf{The SmartInt Approach}}
As we saw, the main challenge we face is handling query constraints as well as projected
attributes that are spread across tables, in the absence of primary key-foreign key dependencies. Broadly speaking, our
approach is to start with a  ``base table" on which most of the query
constraints can be evaluated. The remaining query constraints, i.e.,
those that are over attributes not present in the base table, are {\em
  translated onto the base table} -- i.e., approximated by constraints
over the base table attributes.    The tuples in the base table that
conform to the constraints (both native to the table, and those that
are translated onto it) are the base tuples. After this ``{\bf
  constraint translation}" phase, we enter a ``{\bf tuple expansion}"
phase where the base  tuples are expanded by predicting  values of any
(projected or other) attributes that are not part of the base table. Both the constraint translation and tuple expansion phases are facilitated by inter-attribute correlations called ``approximate functional dependencies" , as well as accompanying value associations that we mine (learn) from samples of the database tables.

\Ignore{ Our approach for addressing these challenges involves two step
starting with a base table containing a subset of query-relevant
attributes, and attempting to ``complete'' the tuples by predicting
the values of the remaining relevant attributes. The
prediction/completion of the tuples is made possible by approximate
functional dependencies, which are automatically mined from samples of
individual tables. The selection of base table itself is influenced by
the confidences of the available AFDs.
Intuitively, the base table
should contain important attributes for whose values cannot be predicted
accurately, but which can help in predicting other attributes. Our
base table selection step formalizes this intuition in terms of the
confidence of the available AFDs.
}

As an illustration of the idea, suppose the following simple
AFDs are mined from our tables (note that the actual mined AFDs can
have multiple attributes on the left hand side):
(1) $S_2: \{ model \} \rightarrow vehicle\_type$, (2) $S_2: \{ model
\} \rightarrow review$, (3) $S_3: \{ model \} \rightarrow cylinders$.
Suppose we start with Table 1 as the base table.
Rule 3 provides us a way to translate the constraint on the number of
cylinders into a constraint on the model (which is present in the
first table). Rules
1 \& 2 provide information on vehicle type and review for a given
model, and hence provide more information in response to the
query. They allow us to expand partial information about the car model
into more complete information about vehicle type, review and
cylinders.  The results using attribute dependencies are shown in Table
~\ref{table:resultsafd} and conform to the constraints and are more
informative compared to other approaches.


\begin{figure}[t]
\centering
\includegraphics[width=3in]{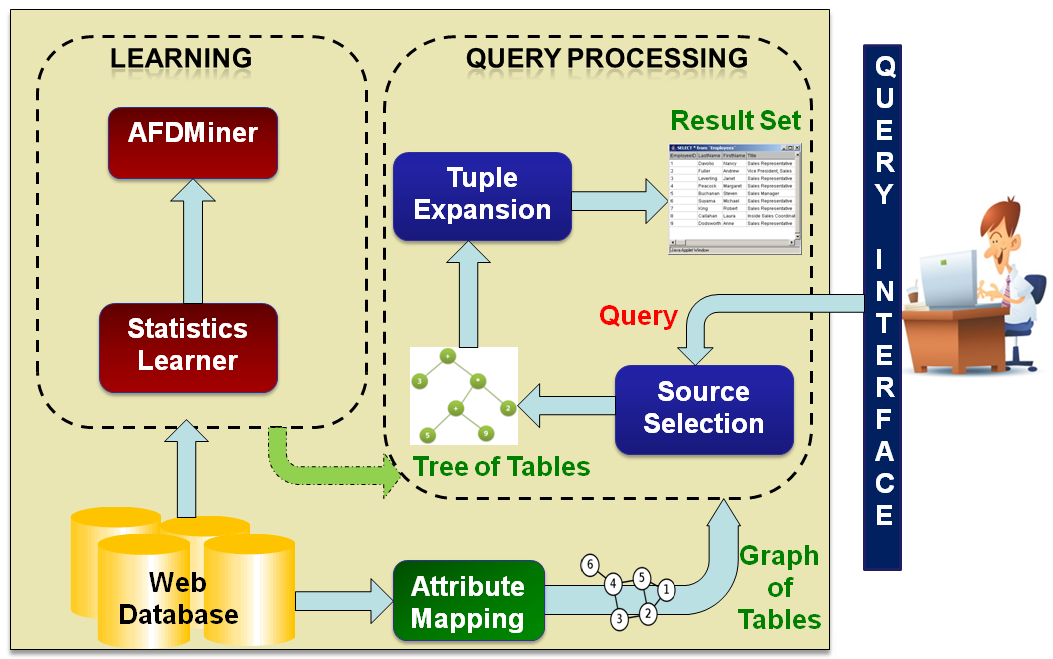}
\caption{Architecture of \smartint\ System}
\label{fig:arch}
\mnvp
\end{figure}

As shown in Figure \ref{fig:arch}, the operation of \smartint\ can thus be understood in terms of (i) mining AFDs and value association statistics from different tables  and (ii) actively using them to propagate constraints and retrieve attributes from other non-joinable tables. Figure \ref{fig:arch} shows the \smartint\ system architecture.
When the user submits a query, the source selector first selects the most relevant `tree' of tables from the available set of tables. The tree of tables provides information about the base table (onto which constraints will be translated), and additional tables from which additional attributes are predicted.
Source selector uses the source statistics mined from the tables to
pick the tree of tables. The tuple expander module operates on the
tree of tables provided by the source selector and then generates the
final result set. Tuple expander first constructs the expanded schema
using the AFDs learned by AFDMiner and then populates the values in
the schema using source statistics.



\smallskip
\noindent{\textbf{Contributions}}
The specific contributions of  \smartint\ system can be summarized as follows:
(i) We have developed a query answering mechanism that utilizes
attribute dependencies to recover entities fragmented over tables,
even  in the absence of primary key--foreign key relations.
 (ii) We have developed a source selection method using novel
relevance metrics that exploit the automatically mined AFDs  to pick the most appropriate set of tables.
(iii) We have developed techniques  to efficiently mine approximate
attribute dependencies. We provide comprehensive experimental results
to evaluate  the effectiveness of \smartint\ as a whole, as well as
its AFD-mining sub-module. Our expeirments are done on data from {\sc
  Google Base}, and show that  \smartint\ is
able to strike a better balance between precision and recall than can
be achieved by relying on single table or employing direct joins.

\smallskip
\noindent{\bf Organization:}
The rest of the paper is organized as follows. Section
\ref{sec:relatedwork} discusses related work about current approaches
for query answering over web databases. Section \ref{sec:defn}
discusses some preliminaries. Section \ref{sec:queryans} proposes a
model for source selection and query answering using attribute
dependencies.Section \ref{sec:learning} provides details about the
methods for learning attribute dependencies. Section
\ref{sec:experiments} presents a comprehensive empirical evaluation of
our approach on data from {\sc Google Base}. Section
\ref{sec:conclusion} provides conclusion and future work. 
A prototype of \smartint\ system has been
demonstrated at ICDE 2010 \cite{smartint-icde}.

\section{Related Work} \label{sec:relatedwork}


\mund{{\bf Data Integration}}
The standard approaches investigated in the database community for the
problem recovering information split across multiple tables is of
course data integration \cite{lenz:di,levy-views,emerac-jiis}.
The approaches
involve defining a global (or mediator) schema that contains all
attributes of relevance, and establishing mappings between the global
schema and the source schemas. This latter step can be done either by
defining global schema relations as views on source relations (called
GAV approach), or defining the source relations as views on the global
schema (called LAV approach). Once such mappings are provided, queries
on the global schema can be reformulated as queries on the source
schemas. While this methodology looks like a hand-in-glove solution to
our problem, its impracticality lies in the fact that it requires
manually or semi-automatically established mappings between global
and source schemas. This
is infeasible in our context where lay users may not even know the set
of available tables, and even if they do, the absence of PK-FK
relations makes establishment of sound and complete
mappings impossible. In contrast, our approach is a fully automated
solution that does not depend on the availability of GAV/LAV
mappings.

\mund{{\bf Entity Identification/Resolution}}
The Entity Identification problem in heterogeneous databases is matching
object instances from different databases that represent same real world
entities. Instance Level Functional Dependencies\cite{lim:entity}
are used to derive missing extended key information for joining the tuples.
Virtual attributes \cite{demichiel:resolving} are found to map databases
in different databases. However, both these approaches require the tables
to have initial key information. Also, it involves manual mappings from the
domain experts to an extent. As opposed to this, \smartint predicts
values for attributes that are not present in the table using mined AFDs.

\mund{{\bf Keyword Search on Databases}}
The entity completion queries handled in \smartint\ are similar in spirit to keyword
queries over databases. This latter has received significant attention
\cite{vagelis:discover,balmin:objectrank,gaurav:banks}. The work on Kite system
extends keyword search to multiple databases as well
\cite{mayassam:keyword}. While Kite doesn't assume that PK-FK
relations are pre-declared, it nevertheless assumes that the columns
corresponding to PK-FK relations are physically present in the
different tables if only under different names. In the context of our
running example, Kite would assume that the model id column is present
in the tables, but not explicitly declared as a PK-FK relation.
Thus Kite focuses on identifying the relevant PK-FK columns
using key discovery techniques (c.f. \cite{TANEFull}).
Their
techniques do not work in the scenarios we consider where the
key columns are simply absent (as we have argued in our motivating
scenario).



\mund{{\bf Handling Incomplete Databases \& Imprecise Queries}} Given a query
involving multiple attributes, \smartint\ starts with a base table containing
a subset of them, and
for each of the tuples in the base table, aims to predict the
remaining query attributes. In this sense it is related to systems such as QPIAD
\cite{QPIAD}. However, unlike  QPIAD which uses AFDs learned from a
single table to complete null-valued tuples, \smartint\  uses AFDs both for translating constraints
onto the base table, and for expanding tuples in the base table by predicting query attributes not in the base table.
 Viewed this way, the critical challenge in
\smartint\ is the
selection of base table, which in turn is based on the confidences of
the mined AFDs (see Section~\ref{sec:queryans}.1).  The constraint translation mechanism used by \smartint\ also has relations to constraint relaxation approaches used by the systems aimed at handling imprecise queries (e.g. \cite{AIMQ}).

\mund{{\bf Learning Attribute Dependencies}} Though rule mining is popular in the database community, the problem of AFD mining is largely under explored. Earlier attempts were made to define AFDs as an approximation to FDs (\cite{CORDS}, \cite{TANEFull}) with few error tuples failing to satisfy the dependency. In these lines, CORDs \cite{CORDS} introduced the notion of Soft-FDs. But, the major shortcoming of their approach is, they are restricted to rules of the type C1$\rightarrow$C2, where C1 and C2 are only singleton sets of attributes. TANE \cite{TANEFull} provides an efficient algorithm to mine FDs and also talks about a variant of the FD-mining algorithm to learn approximate dependencies. But, their approach is restricted to minimal pass rules (Once a dependency of type (X$\rightsquigarrow$Y) is learnt, the search process stops without generating the dependencies of the type (Z$\rightsquigarrow$Y), where X$\subset$Z. Moreover, these techniques are restricted to a single table, but we are interested in learning AFDs from multiple tables and AFDs involving shared attributes. In this paper, we provide a learning technique that treats AFDs as a condensed representation of association rules (and not just approximations to FDs), define appropriate metrics, and develop efficient algorithms to learn all the intra and inter-table dependencies. This unified learning approach has an added advantage of computing all the interesting association rules as well as the AFDs in a single run.


\section{Preliminaries} \label{sec:defn}

Our system assumes that the user does not have knowledge about different tables
in the database and has limited knowledge about attributes he is interested in querying. This is a reasonable assumption, since most web databases
do not expose the tables to the users. So we model the query in the following form where the user just needs to specify the attributes and constraints:
 $\mathcal{Q} = < \bar{A}, \bar{C} >$ where
 $\bar{A}$  are the projected attributes which are of interest to the user and $\bar{C}$ are the
 set of constraints (i.e. attribute-label, value pairs)

Attribute dependencies are represented in the form of approximate functional dependencies.
A \textbf{functional dependency (FD)} is a constraint between two sets of attributes in a relation from a database. Given a relation R, a set of attributes X in R is said to functionally determine another attribute Y, also in R, (written X $\rightarrow$ Y) if and only if each X value is associated with precisely one Y value. Since the real world data is often noisy and incomplete, we use approximate dependencies to represent the attribute dependencies.
An \textbf{Approximate Functional Dependency (AFD)} is an approximate determination of the form $X \rightsquigarrow A$ over relation $R$, which implies that \textbf{attribute set X, known as the determining set}, approximately determines \textbf{A, called the determined attribute.} An AFD is a functional dependency that holds on all but a small fraction of tuples. For example, an AFD $model \rightsquigarrow body\_style$ indicates that the value of a car model usually (but not always) determines the value of $body\_style$.

%


\smallskip

\noindent{\bf Graph of Tables}
The inter-connections between
 different tables in the database are modeled as a graph. Each attribute match is represented as an
undirected edge and any PK-FK relationship is represented as a
directed edge pointing towards the table containing the primary
key.

\section{Query Answering}\label{sec:queryans}

    In this section, we describe our query answering approach. We assume that attribute dependencies are provided upfront for the system. We outline our approach in terms of solutions to challenges identified earlier in Section \ref{sec:intro}:
\\

\mund{\bf 1)  Information distributed across tables needs to be integrated}
            The information needs to be integrated since both answering queries with attributes spanning over multiple tables and providing additional information to the user needs horizontal integration of the tuples across tables. In the absence of PK-FK relationships, performing meaningful joins to integrate data is not feasible (as illustrated in Section \ref{sec:intro}). Instead we start with a `base set of tuples' (from a designated base table chosen by the source selector) and successively expand those tuples horizontally by appending attribute values predicted by the attribute dependencies. This expansion is done recursively until the system cannot chain further or it reconstructs the universal relation. We use attribute determinations along with attribute mappings to identify attributes available in other tables, whose values can be predicted using values of the selected attributes.

\mund{\bf 2) Constraints need to be translated}
            The base table provides a set of tuples, i.e. tuples which conform to the query constraints. Generation of `base set of tuples' requires taking into account constraints on non-base tables. We use attribute mappings and attribute determinations for translating constraints onto the base table. Basically, we need to translate the constraint on a non-base table attribute to a base table attribute through attribute determinations. In the example discussed in Section \ref{sec:intro}, suppose $T_1$ is designated as a base table and $T_3$ is a non-base table  which has an AFD (model $\rightsquigarrow$ vehicle-type). If the query constrains the attribute vehicle-type to be `SUV', then this constraint can be evaluated over the base table, if information about the likelihood of a model being an `SUV' is given. Attribute determinations provide that information.

        Now we explain how these solution approaches are embedded into \smartint\ framework. Query answering mechanism involves two main stages:
        Source Selection and Tuple Expansion. We explain these in
        detail in the next few sections.

\subsection{Source Selection}

        In a realistic setting, data is expected to be scattered across a large number of tables,  and  not all the tables would be equally relevant to the query. Hence, we require a source selection strategy aimed at selecting the top few tables most relevant to the query. Given our model of query answering, where we start with a set of tuples from the base table which are then successively expanded, it makes intuitive sense for tuple expansion to operate over a tree of tables. Therefore source selection aims at returning the most relevant tree of tables over which the Tuple Expander operates.

 Given a user query,  $\mathcal{Q} = < \bar{A}, \bar{C} >$ and a parameter `k' (the number of relevant tables to be retrieved and examined for tuple expansion process), we define source selection as selecting a tree of tables of maximum size
$k$ which  has the highest relevance to the query. The source selection mechanism involves the following steps:
\noindent
1. Generate a set of candidate tables $T_c = \{\mathcal{T} \in T | relevance(\mathcal{T}) \geq threshold\}$.
        This acts as a pruning stage, where tables with low relevance are removed from further consideration.

\noindent
2. Not all tables have a shared attribute. We need to pick a
    connected sub-graph of tables, $G_c$, with highest relevance.

\noindent   
3.  Select the tree with the highest relevance, among all the trees possible in $G_c$.
	This step involves generating and comparing the trees in $G_c$, which can be computationally expensive if $G_c$ is large.	 We heuristically estimate the best tree with the highest relevance to the query among all the trees. The relevance metrics used are explained below.


We will explain how source selection works in the context of the example described in introduction. In order to answer the
query $Q$, \texttt{SELECT make,model WHERE price $<$ \$15000 AND cylinders = `4'}, we can observe that the projected attributes $make$, $model$ and constraint $price < \$15000$ are present in Table \ref{table:schema1} and constraint $cylinders = `4'$ is present in Table \ref{table:schema3}. Given this simple scenario, we can select either Table \ref{table:schema1} or Table \ref{table:schema3} as the base table. If we select Table \ref{table:schema3} as the base table, we should translate the constraint $price < \$15000$ from Table \ref{table:schema1} to Table \ref{table:schema3} using the AFD, $model \rightsquigarrow price$. On the other hand if we designate Table \ref{table:schema1} as base table, we would need to translate the constraint $cylinders = `4'$  from Table \ref{table:schema3} to Table \ref{table:schema1} using the AFD, $model \rightsquigarrow cylinders$.
Intuitively we can observe that the AFD $model \rightsquigarrow cylinders$ generalizes well for a larger number of tuples than $model \rightsquigarrow price$. Source selection tries to select the table which emanates high quality AFDs as the base table and hence yield more precise results.

  Here we discuss the different \textit{relevance functions} employed by the source selection stage:

\mund{\bf Relevance of a table} The relevance of  a table $\cal T$ depends on two factors: (i) the fraction of query-relevant attributes present in the table and--we can view this as ``horizontal relevance" and (ii) the fraction of tuples in the table that are expected to conform to the query --we can view this as ``vertical relevance".  We evaluate relevance as follows:
\vspace*{-0.1in}
	$$ relevance(\mathcal{T},q) \approx  \frac {|A_\mathcal{T} \cap {\bar{A}}|} {|{\bar{A}}|} * Pr_\mathcal{T}(\bar{C})*tupleCount_\mathcal{T}$$
	where the first factor is measuring the horizontal relevance and the other two estimate the vertical relevance. Specifically, $Pr_\mathcal{T}(\bar{C})$ is the probability that a random tuple from $\mathcal{T}$ conforms to constraints $\bar{C}$,
	$tupleCount_\mathcal{T}$ is the number of tuples in $\mathcal{T}$, and $A_\mathcal{T}$ is the set of attributes in $\mathcal{T}$.\footnote{Presently we give equal weight to all the attributes in the system, this can be generalized to account for attributes with
different levels of importance.}

\mund{\bf Relevance of a tree}
    While selecting the tree of relevant tables, the source selection stage needs to estimate the relevance of tree.
    The relevance of tree takes into account the confidence of AFDs emanating out of the
   table.
	Relevance of a tree $T_r$ rooted at table $\mathcal{T}$ w.r.t query $Q <\bar{A},\bar{C}>$ can be expressed as: 
    $relevance(T_r, q) = relevance(\mathcal{T}, q) + 
 \sum_{a \in \bar{A} - A_b} pred\_accuracy(a) $
    where $A_b$ are the set of attributes present in the base table,
    and 
 $pred\_accuracy(a)$ gives the accuracy with which the attribute $a$ can be predicted.
    When the attribute is in the neighboring table it is equal to the
    confidence of AFD and when its not in the immediate neighbor its
    calculated the same way as in AFD chaining (Explained in Section \ref{sec:learning}).



    The above relevance functions rely on the conformance probability $P_\mathcal{T}(C) = \Pi_i P_\mathcal{T}(C_i)$.
	$P_\mathcal{T}(C_i)$ denotes the probability that a random tuple from $\mathcal{T}$ conforms to the constraint $C_i$ (of the form $X = v$), and is estimated as:
	\begin{itemize}
	\item $P_\mathcal{T}(C_i) = P_\mathcal{T}(X = v), if X \in A_\mathcal{T}$,
	where $A_\mathcal{T}$ is the set of attributes in $\mathcal{T}$
	\item $P_\mathcal{T}(C_i) = \sum_i P_\mathcal{T}(Y = v_i) * P_\mathcal{T'}(X = v | Z = v_i)$,
	if $\mathcal{T} : Y = \mathcal{T'} : Z$, i.e. $\mathcal{T}$'s
        neighboring table $\mathcal{T'}$ provides attribute X. (These probabilities are learnt as source statistics.)
	\item $P_\mathcal{T}(C_i) = \epsilon $ (small non-zero probability), otherwise
	\end{itemize}

\begin{algorithm}[ht]
\caption{\label{src-select}Source Selection}
\mvp
\begin{algorithmic}[1]
\REQUIRE  Query q, Threshold $\tau$, Number of tables k, Set of AFDs $\bar{A}$
\STATE $T_c$ = $\{\emptyset\}$
\FORALL {table $\mathcal{T}$ in T }
\IF { relevance($\mathcal{T}$, q) $\geq$ $\tau$ }
\STATE add $\mathcal{T}$ to $T_c$
\ENDIF
\ENDFOR
\STATE $G_c$ := Set of connected graphs over $T_c$ up to size k
\STATE $Trees$ = $\{\emptyset\}$
\FORALL {$g \in G_c$}
\STATE $Trees_g$ = Set of trees from graph g
\STATE add $Trees_g$ to $Trees$
\ENDFOR
\STATE $tree_{sel} = \arg\max_{tree \in Trees} relevance(tree, q)$
\STATE {\bf return} $tree_{sel}$
\end{algorithmic}
\end{algorithm}

   In this section we explained the source selection mechanism. We discuss how the tuple expansion mechanism
    answers the query from the selected sources in the next section.

\subsection{Tuple Expansion}

Source selection module gives a tree of tables which is most relevant to the query. Tuple expansion operates on the tree of tables given by that module. One of the key contributions of our work is returning the result tuples with schema as close to the universal relation as possible. We need to first construct the schema for the final result set and then populate tuples that correspond to that particular schema from other tables. These steps are described in detail in the sections that follow.


\subsubsection{Constructing the Schema}
\begin{figure}
\centering
\includegraphics[width=3in]{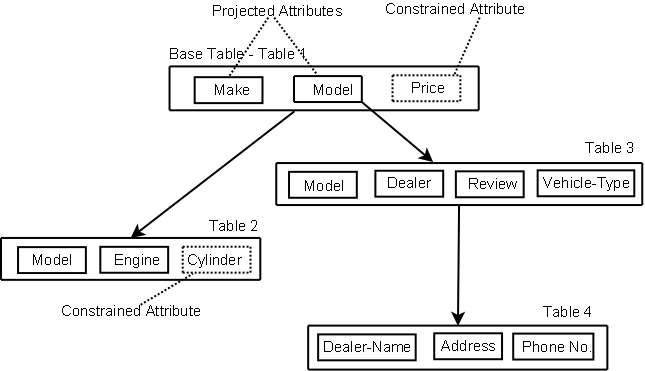}
\caption{Expanded attribute tree for the query}
\mvp
\label{fig:tupexp}
\end{figure}

 One important aspect of tuple expansion is that it is a hierarchical
 expansion. The schema grows in the form of a tree because attributes
 retrieved from other tables are relevant only to the determining
 attribute(s)(refer to the definition of AFD in Section
 \ref{sec:defn}). This module returns a hierarchical list of
 attributes, {\it AttrbTree}, rather than a flat list. This is more
 clearly illustrated  by the attribute tree generated for query
 discussed in Section \ref{sec:intro} shown in Figure
 \ref{fig:tupexp}. The base table ($T_1$) contains attributes {\it
   Make, Model, Price}. Tables $T_1$, $T_2$ and $T_3$ share  the
 attribute {\it Model}. In table $T_2$, we have the AFDs {\it Model
   $\rightsquigarrow$ Cylinders} and {\it Model $\rightsquigarrow$
   Engine}. These two determined attributes are added to the base
 answer set, but these are only relevant to the attribute `model', so
 they form a branch under the attribute `model'. Similarly, {\it
   review, dealer} and {\it vehicle type} form another branch under
 `Model'. In the next level, $T_3$ and $T_4$ share `dealer-name'
 attribute. `Dealer-Name' is a key in $T_4$, therefore all the
 attributes in $T_4$ (`dealer-address', `phone-number' etc) are
 attached to the \emph{AttrbTree}. The final attribute tree is shown
 in the Figure ~\ref{fig:tupexp}.

\subsubsection{Populating the Tuples}

The root of the selected tree of tables given by the source selection is designated as the {\it base table}. Once the attribute hierarchy is constructed, the system generates a `base set' of tuples from the base table which form the `seed' answers.
We refer to this base set as the \textit{most likely tuples} in the base table which conform to the constraints mentioned in the query. We call them `most likely' tuples because when constraints are specified on one of the children of the base table, we propagate constraints from child to base table. But since we have approximate dependencies between attributes, the translated constraints do not always hold on the base set. To clearly illustrate this, let us revisit the example of \textit{Vehicle} domain from Section \ref{sec:intro}. We assume that Table \ref{table:schema1} has been designated as the base table. The constraint $price < \$15000$ is local for the base table and hence each tuple can be readily evaluated for conformance. The constraint
\emph{cylinders = `4'}, on the other hand, is over Table \ref{table:schema3} and needs to be translated on to the base table. Notice that these two tables share the attribute `model' and this attribute can approximately determine \textit{cylinder} in Table \ref{table:schema3} ( $model \rightsquigarrow Cylinders$ ). ($model \rightsquigarrow Cylinders$) implies that the likelihood of a model having certain number of cylinders can be estimated, which can be used to estimate the probability that a tuple in Table 1 would conform to the constraint $Cylinders = `4'$. We can see that model `Civic' is more likely to be in the base set than `Accord' or `Camry'.

Once the base tuple set has been generated, each of those tuples are expanded horizontally by predicting the values for the attributes pulled from children tables. Given a tuple from the base set, all the children tables (to the base table) are looked up for determined attributes, and the most likely value is used to expand the tuple. Further, values picked from the children tables are used to pick determined attributes from their children tables and so on. In this way, the base tuple set provided by the root table is expanded using the learned value dependencies from child tables.

In tuple expansion, if the number of shared attributes between tables is greater than one, getting the associated
values from other tables would be an interesting challenge. For instance, in our running example, Table \ref{table:schema1} also had the year attribute and Table \ref{table:schema2} is selected as the base table. We need to predict  the value of price from Table \ref{table:schema1}. If we consider both Model and Year to predict the price, results would be more accurate, but we do not have the values of all combinations of Model
and Year in Table \ref{table:schema1} to predict the price. However, if we just use Model to predict the price, the precision might go down. Another interesting scenario where taking multiple attributes might not boost the prediction accuracy is the following: {\it Model, Number\_tires $\rightsquigarrow$ Price} is no better than {\it Model $\rightsquigarrow$ Price}. In order to counter this problem, we propose a \textit{fall back} approach of the AFDs to ensure high precision and recall.

This method can be formally described as this: If $\mathcal{X}$ is the set of shared attributes
between two tables $T_1$ and $T_2$, where $T_1$ is the base table and $T_2$ is the child table. We need to predict the values of attribute $Y$ from $T_2$ and populate the result attribute tree. If the size of $\mathcal{X}$ is equal to $n$ ($n \geq 1$), we would first start with AFDs having $n$ attributes in determining set and `significantly higher' confidence than any of their AFDs. We need `significantly higher' confidence because if the additional attributes do not boost the confidence much, they will not increase the accuracy of prediction as well.
If the AFDs do not find matching values between two tables to predict values, we `fall back' to the AFDs with smaller determining set. We do this until we would be able to predict the value from the other table.  Algorithm \ref{algo:tupex} describes it.

\begin{algorithm}[ht]
\caption{\label{algo:tupex}Tuple Expansion}
\mvp
\begin{algorithmic}[1]
\REQUIRE  Source-table-tree $\mathcal{S}_t$; Result-attribute-tree $\mathcal{A}_t$, Set of AFDs $\bar{A}$
\STATE $R$ := $\{\emptyset\}$  \COMMENT{Initializing the result set with schema $\mathcal{A}_t$ }
\STATE $b$ := $Root(\mathcal{S}_t)$ \COMMENT{Setting the base table}
\STATE Translate the constraints onto base table
\STATE Populate all the attributes in level 0 of $\mathcal{A}_t$ from $b$
\FORALL {child $c$ in $\mathcal{A}_t$}
\IF {$b$ and $c$ share $n$ attributes}
\STATE ${fd}$ = AFDs with $n$ attrbs in detSet
\WHILE {$n > 0$}
\IF {$c$ has the specified combination }
\STATE Populate $R$ using predicted values using ${fd}$ from $c$
\STATE {\bf break}
\ENDIF
\STATE ${fd}$ = Pick AFDs with $n-1$ attributes in detSet
\ENDWHILE
\ENDIF
\ENDFOR
\STATE {\bf return} Result Set $R$
\end{algorithmic}
\end{algorithm}


\section{Learning Attribute Dependencies}\label{sec:learning}

We have seen in the previous section how attribute dependencies within and across tables help us in query answering by discovering related attributes from other tables. But it is highly unlikely that these dependencies will be provided up front by autonomous web sources. In fact, in most cases the dependencies are not apparent or easily identifiable. We need an automated learning approach to mine these dependencies.

As we have seen the Section \ref{sec:queryans}, we extensively use both attribute-level dependencies (AFDs) and value level dependencies. The value level dependencies are nothing but \textit{association rules}. The notion of mining AFDs as condensed representations of association rules is discussed in detail in \cite{ak:thesis}. Our work adapts the same notion, since it helps us in learning dependencies both at attribute and value level.

The following sections describe how rules are mined within the table and how they are propagated across tables.

\subsection{Intra-table learning}
In this subsection we describe the process of learning AFDs from a
single table. It is easy to see that the number of possible AFDs in a
database table is exponential to the number of attributes in it, thus
AFD mining is in general expensive. But, only few of these AFDs are
useful to us. To capture this, we define two metrics \ConfAFD\ and
\SuppInfo\ for an AFD, and focus on AFDs that have high \ConfAFD\ and
low \SuppInfo\ values.

\subsubsection{Confidence }

If an Association rule is of the form $(\alpha \rightsquigarrow
\beta)$, it means that if we find all of $\alpha$ in a row, then we
have a good chance of finding $\beta$. The probability of finding
$\beta$ for us to accept this rule is called the confidence of the
rule. Confidence denotes the conditional probability of head given the
body of the rule.

Generalizing to AFDs, the confidence of an AFD 
should similarly  denote the chance of finding the value for the
dependent attribute, given the values of the attributes in the
determining set. We define \ConfAFD\ in terms of the confidences of
the underlying association rules. Specifically, we define it in terms
of picking the best association rule for every distinct
value-combination of the body of the association rules. For example,
if there are two association rules (Honda $\rightsquigarrow$ Accord)
and (Honda $\rightsquigarrow$ Civic), given Honda, the probability of
occurrence of Accord is greater than the probability of occurrence of
Civic. Thus, (Honda $\rightsquigarrow$ Accord) is the best association
rule, for (Make = Honda) as the body.  \vspace*{-0.1in}
\begin{eqnarray}
\emph{\ConfAFD\ (X$\rightsquigarrow$ $A$)} &=& \sum_{x}^{N^{'}} \arg\max_{y \in [1, N_{j}]} (\emph{support}\left(\alpha_{x}\right) \times \nonumber \\
& & Confidence( \alpha_{x} \rightsquigarrow \beta_{y}))
\nonumber
\end{eqnarray}
Here, $N^{'}$ denotes the number of distinct values for the determining set X in the relation.
This can also be written as,
\vspace*{-0.1in}
\begin{equation}
\emph{Confidence (X$\rightsquigarrow$ A)}= \sum_{x}^{N^{'}} \arg\max_{y \in [1, N_{j}]} (\emph{support}\left(\alpha_{x}\right) \rightsquigarrow \beta_{y})
\nonumber
\end{equation}

\begin{table}[!t]
\centering

    \caption{Fragment of a Car Database}
\mvp
    \label{carDBTable}

\begin{small}
\begin{tabular}{|c|c|c|c|c|}
  \hline
  \bf{ID} & \bf{Make} & \bf{Model} & \bf{Year} & \bf{Body Style} \\\hline
  1 & Honda & Accord & 2001 & Sedan \\\hline
  2 & Honda & Accord & 2002 & Sedan \\\hline
  3 & Honda & Accord & 2005 & Coupe \\\hline
  4 & Honda & Civic & 2003 & Coupe \\\hline
  5 & Honda & Civic & 1999 & Sedan \\\hline
  6 & Toyota & Sequoia & 2007 & SUV \\\hline
  7 & Toyota & Camry & 2001 & Sedan \\\hline
  8 & Toyota & Camry & 2002 & Sedan \\\hline
\end{tabular}
\end{small}

\end{table}

Example:
For the database relation displayed in table \ref{carDBTable}, Confidence of the AFD $(Make\rightsquigarrow Model)$ = Support $(Make:Honda\rightsquigarrow Model:Accord)$ + Support $(Make:Toyota\rightsquigarrow Model:Camry)$ = $\frac{3}{8}$ + $\frac{2}{8}$ = $\frac{5}{8}$.

\subsubsection{Specificity-based Pruning}
 The distribution of values for the determining set is an important measure to judge the ``usefulness"
of an AFD. For an AFD $X
\rightsquigarrow A$, the fewer distinct values  of $X$ and the more
tuples in the database that have the same value, potentially the
more relevant possible answers can be retrieved through each query, and thus a better recall. To quantify this, we first
define  the {\em support of a value} $\alpha_{i}$ of an attribute
set $X$, $support(\alpha_{i})$, as the occurrence frequency of value
$\alpha_{i}$ in the training set. The support is defined as 
$support (\alpha_{i}) =  count(\alpha_{i})/N,$
where $N$ is the number of tuples in the training set.

Now we measure how the values of an attribute set $X$ are
distributed using \SuppInfo. \SuppInfo\ is defined as the
information entropy of the set of all possible values of attribute
set $X$:
 \{$\alpha_{1}$, $\alpha_{2}$, \ldots, $\alpha_{m}$ \},
normalized by the maximal possible  entropy (which is achieved when
$X$ is a key). Thus, \SuppInfo\ is a value that lies between 0 and
1.

\vspace*{-0.2in}
\begin{eqnarray}
 \emph{\SuppInfo\ ($X$)}
&=& \frac{- \sum_{1}^{m}
support(\alpha_{i})\times\log_{2}(support(\alpha_{i}))}{\log_{2}(\emph{N})}
\nonumber
\end{eqnarray}

When there is only one possible value of $X$, then this value has
the maximum support and is the least specific, thus we have
\SuppInfo\ equals to 0.
When all values of $X$ are distinct, each value has the minimum
support and is most specific. In fact, $X$ is a key in this case and
has \SuppInfo\ equal to 1.

Now we overload the concept of \SuppInfo\ on AFDs. The \SuppInfo\ of
an AFD is defined as the \SuppInfo\ of its determining set. i.e. 
$\emph{\SuppInfo\ ($X \rightsquigarrow A$)}  = \emph{\SuppInfo\ ($X$)}$.
The lower \SuppInfo\  of an AFD, potentially the more  relevant
possible answers can be retrieved using the rewritten queries
generated by this AFD, and thus a higher recall for a given number
of rewritten queries.

Intuitively, \SuppInfo\ increases when the number of distinct
values for a set of attributes increases. Consider two attribute sets $X$ and $Y$ such that Y$\supset$X. Since $Y$ has more
attributes than $X$, the number of distinct values of $Y$ is no less
than that of $X$, \SuppInfo\ (Y) is no less than  \SuppInfo\ ($X$).

%

\begin{definition}[Monotonicity of \SuppInfo\ ]
For any two attribute sets $X$ and $Y$ such that Y$\supset$X, \SuppInfo\ (Y) $\geq$ \SuppInfo\
($X$). Thus, adding more attributes to the attribute set $X$ can only increase the \SuppInfo\ of $X$. Hence, \SuppInfo\ is monotonically increasing w.r.t increase in the number of attributes.
\end{definition}

This property is exploited in pruning the AFDs during the mining, by eliminating the search space of rules with \SuppInfo\ less than the given threshold. 

Algorithms for mining AFDs face two costs: the combinatorial cost of searching the rule space and the cost of scanning the data to calculate the required metrics for the rules. In query processing the AFDs which we are mostly interested are the ones with the shared attributes in determining set of the rule. If $X \rightsquigarrow A$ is an AFD, we are interested in rules where $X \in S$, where $S$ is the set of shared attributes between two tables. Since number of such attributes is typically small, we can use this as one of the heuristics to prune away irrelevant rules.

\subsubsection{AFDMiner algorithm}

The problem of mining AFDs can be formally defined as follows:
Given a database relation \textbf{r}, and user-specified thresholds $\emph{minConf}$ (minimum confidence) and \emph{max}\SuppInfo\ (maximum \SuppInfo\ ), generate all the Approximate Functional Dependencies (AFDs) of the form ($X\rightsquigarrow A$) from \textbf{$r$} for which \ConfAFD\ (X$\rightsquigarrow A) \geq minConf$ and \SuppInfo\ (X) $\leq $ \emph{max}\SuppInfo\


To find all dependencies according to the definition above, we search through the space of non-trivial dependencies and test the validity of each dependency. We follow a breadth first search strategy and perform a level-wise search in the lattice of attributes, for all the required AFDs. Bottom-up search in the lattice starts with singleton sets and proceeds upwards level-wise in the lattice, searching bigger sets. For AFDs, the level-wise bottom-up algorithm has a powerful mechanism for pruning the search space, especially the pruning based on \SuppInfo\ .

Search starts from singleton sets of attributes and works its way to larger attribute sets through the set containment lattice level by level. When the algorithm is processing a set X, it tests AFDs of the form $X \setminus {A}\rightsquigarrow A$, where A $\in$ X.

\begin{algorithm}[ht]
\caption{\label{mainAlgo}AFDMiner: Levelwise search of dependencies}
\mvp
\begin{algorithmic}[1]
\STATE \emph{$L_{0}$} := $\{\emptyset\}$
\STATE \emph{$L_{1}$} := $\{\{\emph{A}\} \mid \emph{A} \in \emph{R} \} $
\STATE $\ell$ := 1
\WHILE{\emph{$L_{\ell} \neq \emptyset$}}
\STATE ComputeDependenciesAtALevel(\emph{$L_{\ell}$})
\STATE PRUNE(\emph{$L_{\ell}$})
\STATE \emph{$L_{\ell+1}$} := GenerateNextLevel(\emph{$L_{\ell}$})
\STATE \emph{\emph{$\ell$}} := \emph{\emph{$\ell$}} + 1
\ENDWHILE
\end{algorithmic}
\end{algorithm}

Algorithm \ref{mainAlgo} briefly presents the main $AFDMiner$ algorithm.
In it, GenerateNextLevel computes the level $L_{\ell+1}$ from $L_{\ell}$. The level $L_{\ell+1}$ will contain only those attribute sets of size $\ell+1$ which have their subsets of size $\ell$ in $L_{\ell}$. (ComputeDependenciesAtALevel(\emph{$L_{\ell}$})) computes all the AFDs that hold true at the given level of the lattice. In this process, it computes the confidence of eah association rule constituting the AFDs.
PRUNE(\emph{$L_{\ell}$}) implements the pruning strategies and prunes
the search space of AFDs. It computes the \SuppInfo\ of each rule, and
if it is less than the specified threshold, eliminates all the rules 
whose determining sets are supersets of it.

\subsection{Learning source statistics}  \label{sec:learning_source_stats}
\mund{\textbf{Storing association rules}}
The probabilities which we used extensively in the query answering phase are nothing but the confidence of the association rules. So we store all the association rules mined during the process of AFD mining (specifically, in ComputeDependenciesAtALevel(\emph{$L_{\ell}$}))) and use them at query time. This saves us the additional cost of having to compute the association rules separately by traversing the whole lattice again.

Here we describe the value level source statistics gathered by the system, which are employed by the
query answering module for constraint propagation and attribute value prediction. As mentioned earlier, AFD mining involves mining the underlying association rules. During association rule mining, following statistics are gathered from each source table $\mathcal{T}$:
%
(i)  \textbf{$P_\mathcal{T}(X = x_i)$:} Prior probabilities of distinct values for each attribute X in $A_\mathcal{T}$
(ii) \textbf{$P_\mathcal{T}(X = x_i | Y = y_j)$:} Conditional probabilities for distinct values of each attribute X conditioned on those of attribute Y in $A_\mathcal{T}$.
Recall that this is nothing but the confidence of an association rule. Only the shared attributes are used as evidence variables, since value prediction and constraint propagation can only be performed across shared attributes.

 \begin{figure}
 \centering
 \includegraphics[width=3in]{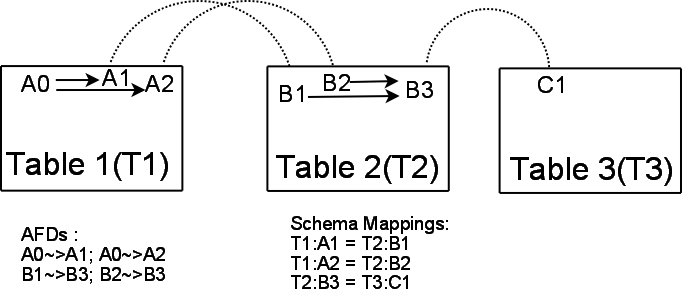}
 \caption{Inter-table Learning}
 \label{fig:intertable}
 \end{figure}


\subsection{Inter-table Chaining}

After learning the AFDs within a table, we need to use them to derive inclusion dependencies
which are used in query answering phase. In order to combine AFDs
from different tables, we need anchor points. These anchor points are
provided by the attribute mappings across tables, so we extend our
attribute dependencies using them. When two AFDs between neighboring tables are combined, the resultant AFD would have a confidence equal to the product of the two confidences.

 But when we are combining dependencies between tables which are not directly connected, we need to consider all the possibilities. Let us consider the scenario in Figure \ref{fig:intertable}, with three tables
 $T_1, T_2$ and $T_3$. $T_1$ and $T_2$ have mappings between attributes $A_1 - B_1$ and $A_2 - B_2$. Similarly $T_2$ and $T_3$ have mapping between $B_3 - C_1$. If we want to get the most likely value of $C_1$ for $A_0$,
 we have more than one chaining to consider. We need to consider the confidences of AFDs, $A_0 \rightsquigarrow A_1$, $A_0 \rightsquigarrow A_2$ as well as the confidences of AFDs, $B_1 \rightsquigarrow B_3$, $B_2 \rightsquigarrow B_3$. We cannot greedily pick the AFD with higher confidence in either $T_1$ or $T_2$. We need to pick a combination of the AFDs which have higher \textit{cumulative} confidence.
%

\section{Experimental Evaluation}\label{sec:experiments}

\begin{figure}
\centering
\includegraphics[width=2.9in]{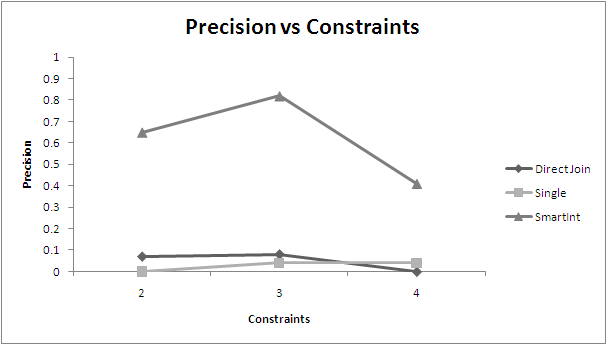}
\caption{Precision {\em vs.} Number of Constraints}
\mnvp
\label{fig:p-const}
\vspace*{-0.1in}
\end{figure}

\Ignore{
In this section, we describe the implementation and an empirical evaluation of our system \smartint\ for query processing over multiple tables and learning attribute dependencies. Our prototype works on a local copy of the web databases for both efficiency as well as due to access restrictions for many of the Web databases. We implemented our system in Java and used MySQL database to store the tables.}

A prototype of \smartint\ system, as described in this paper, has been
implemented. The prototype supports automatic mining of approximate
functional dependencies and value associations in an off-line phase.
It also ranks the answer tuples it returns in terms of the overall
confidence associated with each tuple. The prototype system has been
demonstrated at ICDE 2010 \cite{smartint-icde}.

\begin{figure}
\centering
\includegraphics[width=2.9in]{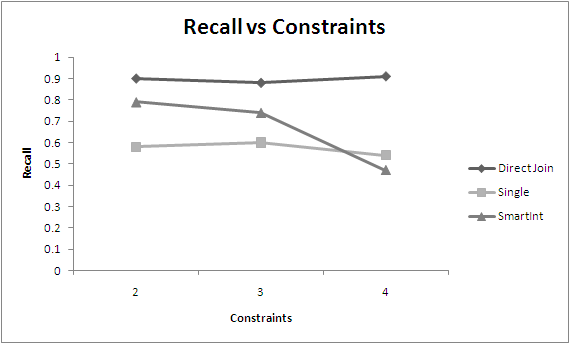}
\caption{Recall {\em vs} Number of Constraints}
\mvp
\label{fig:r-const}
\vspace*{-0.1in}
\end{figure}

\begin{figure}
\centering
\includegraphics[width=2.9in]{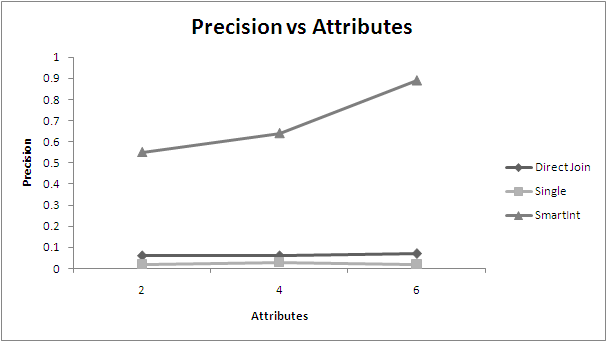}
\caption{Precision {\em vs} Number of Attributes}
\mnvp
\label{fig:p-attrb}
\end{figure}


Our intent is to evaluate the effectiveness of \smartint\ in terms of precision and recall measures.
The following explains how precision and recall measures are computed to take into account the fact that
\smartint's answers  can differ from ground truth(provided by the
master table) both in terms of how many answers it returns and how correct
and complete each answer is.

{\em Correctness of a tuple} ($cr_t$) is defined as the fraction
of its attribute values that are correct.  {\em Precision of the
  result set} ($P_{rs}$) is defined as the average correctness of the
tuples in the result set.
Similarly, {\em completeness of a tuple}($cp_t$) is defined as the
fraction of the attributes of the real tuple that were returned.
%
%
{\em Recall of the result set} ($R_{rs}$) is defined as the
average completeness of the tuples in the result set. 

\mund{Experimental Setup} \label{app:exp}
To evaluate the \smartint\ system, we used Vehicles database. We used around 350,000 records probed from {\bf Google Base} for the experiments. We created a \emph{master table} with 18 attributes. We divided this \textit{master table} into multiple child tables with overlapping attributes. This helps us in evaluating the returned `result set' with respect to the results from master table and establish how our approach compares with the ground truth.We have divided the master table into 5 different tables with the following schema

\textit{\begin{small}\begin{itemize}
\item Vehicles\_Japanese: (condition, price\_type, engine, model, VIN, vehicle\_type, payment,door\_count, mileage, price, color , body\_style, make)
\item Vehicles\_Chevrolet: (condition, year, price, model, VIN, payment, mileage, price, color, make),
\item Vehicles\_Chevrolet\_Extra: (Model, Door Count, Type,  Engine)
\item Vehicles\_Rest: (condition, year, price ,model, VIN, payment, mileage, price, color, make)
\item Vehicles\_Rest\_Extra: (Engine, Model, Vehicle Type, door count, body style)
\end{itemize}\end{small}}

\begin{figure}
\centering
\includegraphics[width=2.9in]{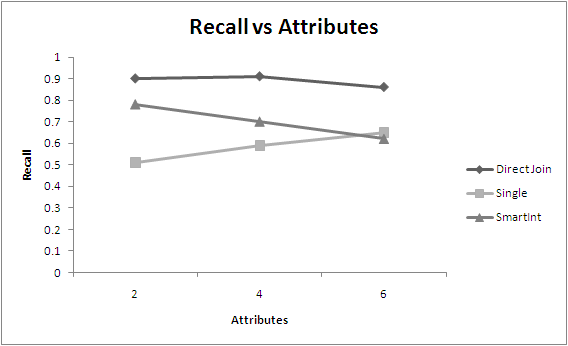}
\caption{Recall {\em vs} Number of Attributes}
\mnvp
\label{fig:r-attrb}
\end{figure}

The following (implicit) attribute overlaps were present among the fragmented tables.
\textit{\begin{small}\begin{itemize}
\item {Vehicles\_Chevrolet:Model $\leftrightarrow$ Vehicles\_Rest:Model}
\item {Vehicles\_Chevrolet:Year $\leftrightarrow$ Vehicles\_Rest:Year}
\item {Vehicles\_Rest:Year $\leftrightarrow$ Vehicles\_Rest\_Extra:Year}
\item {Vehicles\_Chevrolet\_Extra:Model $\leftrightarrow$ Vehicles\_Rest\_Extra:Model}.
\end{itemize}\end{small}}

The following are the input parameters which are
changed: (1) Number of Attributes and (2) Number of Constraints.
We measured the value of precision and recall by taking the average of the values for different queries. While measuring the value for a particular value of a parameter we varied the other parameter. While we are measuring precision for `Number of attributes = 2', we posed queries to the system with `Number of constraints = 2, 3 and 4' and took the average of all these values and plotted them. Similarly, we varied the `Number of attributes' while we are measuring the Precision for each value of `Number of constraints'. The same process is repeated for measuring the recall as well.

\mund{Comparison with `Single table' and `Direct join' approaches }

In this section, we compare the accuracy of \smartint\ with `Single
table' and `Direct join' approach which we discussed in
\ref{sec:intro} and analyze them.  Recall that in the single table
approach, results are retrieved from a single table which has maximum
number of attributes/constraints mentioned in the query mapped on it.
The direct join approach involves joining the tables based on the
shared attributes. As explained in the introduction, the latter
approach tends to generate spurious entities, while the former also
fails to draw together the connected information about the entity.

\begin{figure}
\centering
\includegraphics[width=2.9in]{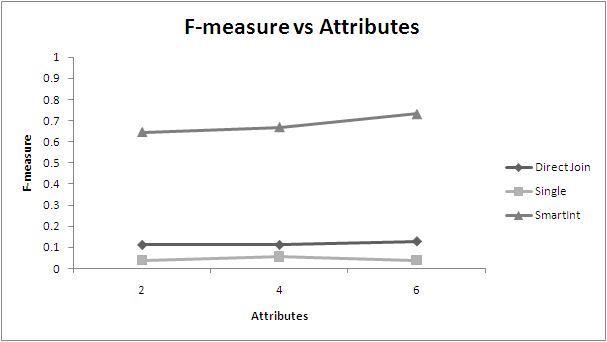}
\caption{F-measure {\em vs} Number of Attributes}
\mnvp
\label{fig:fm-attrb}
\end{figure}

In the simple case of queries mapping on to a single table, the precision and recall values are independent of
attribute dependencies, since query answering does not involve constraint propagation or tuple expansion through
attribute value prediction.

\begin{figure}
\centering
\includegraphics[width=2.9in]{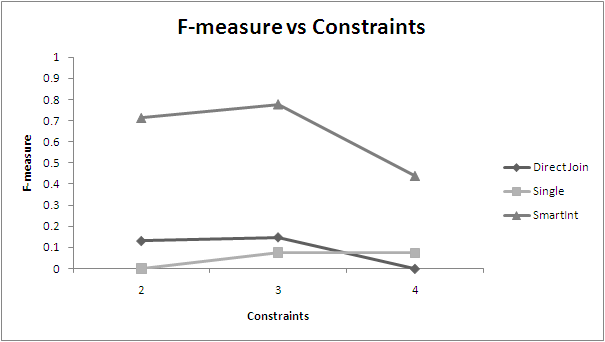}
\caption{F-measure {\em vs} Number of Constraints}
\mnvp
\label{fig:fm-const}
\end{figure}

In cases where queries span multiple tables, some of the attribute values have to be predicted and constraints have to be propagated across tables.
Availability of attribute dependency information allows accurate prediction of attributes values and hence boosts precision.
As shown in Figures \ref{fig:p-const} and \ref{fig:p-attrb}, our approach scored over the other two in precision. Direct join approach, in absence of primary-foreign key relationships,
ends up generating non-existent tuples through replication, which severely compromises the precision. In cases where query constraints span over multiple tables,
single table approach ends up dropping all the constraints except the ones mapped on to the selected best table. This again results in low precision.

In terms of recall (Figures \ref{fig:r-const} and \ref{fig:r-attrb}), performance is dominated by the direct join approach, which is not surprising. Since direct join combines partial answers from selected tables,
the resulting tuple set contains most of the real answers, subject to completeness of individual tables. Single table approach, despite dropping constraints, performs
poorly on recall. The selected table does not cover all the query attributes, and hence answer tuples are low on completeness, which affects recall.

When accurate attribute dependencies are available, our approach processes the distributed query constraints effectively and
hence keeps the precision fairly high. At the same time, it performs chaining across tables to improve the recall.
Figures \ref{fig:fm-attrb} and \ref{fig:fm-const} show that our approach scores higher on F-measure, hence suggesting that
it achieves a better balance between precision and recall.

\mund{Comparison with multiple join paths}

\begin{figure}
\centering
\includegraphics[width=2.9in]{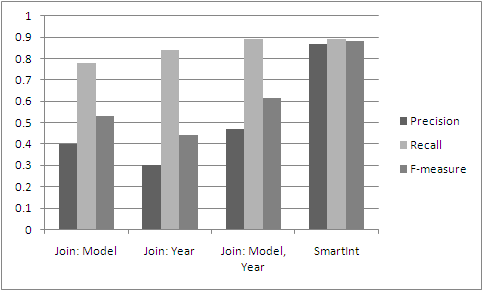}
\caption{\smartint\ {\em vs} Multiple join paths}
\mvp
\label{fig:multi-join}
\vspace*{-0.1in}
\end{figure}

In the previous evaluation the data model had one shared attribute between the tables, but there can be multiple shared attributes between the tables. In such scenarios, direct join can be done based on any combination of the shared attributes. Unless one of the attribute happens to be a key column the precision of the joins is low. In order to illustrate this, we considered the data model with more than one shared attribute and measured the precision and recall for all the possible join paths between the tables. The experimental results (See Figure \ref{fig:multi-join}) show that \smartint\ had higher F-measure than all possible join paths.

\mund{Tradeoffs in number vs. completeness of the answers}
\begin{figure}
\centering
\includegraphics[width=2.9in]{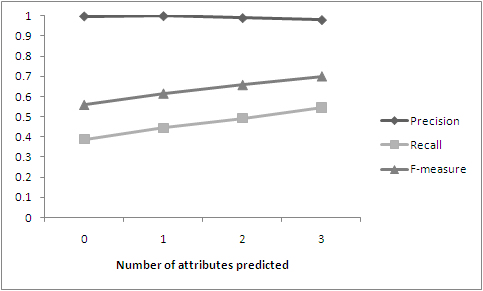}
\caption{Precision, Recall and F-measure {\em vs} tuple width}
\mnvp
\label{fig:topkw}
\end{figure}

Normal query processing systems are only
concerned about retrieving top-k results since the width(number of
attributes) of the tuple is fixed. But \smartint\ chains across the
tables to increase the extent of completion of the entity. This poses
an interesting tradeoff: In a given time, the system can retrieve more
tuples with less width or fewer  tuples with more width. In
addition to this, if user is only interested high confidence answers,
each tuple can expand to variable width to give out high precision
result set. We analyze how precision and recall varies with $w$
(the number of attributes to be shown). The Figure \ref{fig:topkw} shows
how precision, recall and F-measure varies as more number of
attributes are predicted for a specific result set (the query
constraints are make=`BMW' and year=`2003'). In scenarios, when
\smartint\ has to deal with infinite width tuples, F-measure can be
used to guide \smartint\ when to stop expanding.

\mund{Learning Time (AFDMiner)}

\begin{figure}
\centering
\includegraphics[width=2.9in]{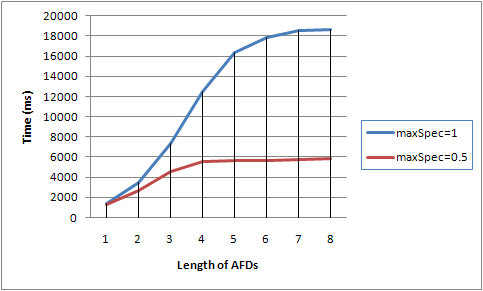}
\caption{Time taken by AFDMiner {\em vs} Length of AFD}
\mnvp
\label{fig:afdlength}
\end{figure}

\begin{figure}
\centering
\includegraphics[width=2.9in]{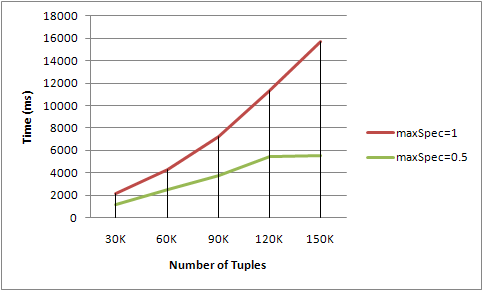}
\caption{Time taken by AFDMiner {\em vs} No. of tuples}
\mnvp
\label{fig:learningtime}
\end{figure}

We invoke AFDMiner to learn the association rules and the AFDs. But this is done offline before query processing starts. So learning time usually does not directly affect the performance of the system.
Nevertheless, the current implementation of AFDMiner uses several
optimizations and data preprocessing  to keep learning time  low. In
fact, {\em AFDMiner takes only about 4 seconds for mining the rules
  used in the current experimental setup}. Figure
\ref{fig:learningtime} and \ref{fig:afdlength} show the comparison
between the time taken for AFDMiner with specificity threshold set to
0.5 and 1, with varying tuplesize and the length of the AFD
respectively.\footnote{At first blush, pruning highly specific AFDs
  seems to hurt the precision, but in the current set of experiements
  \SuppInfo\ based pruning reduced the total running time and did not
  effect the accuracy.} We see that \SuppInfo\ metric results in
faster learning times.
For a detailed experimentation on AFDMiner, refer to \cite{ak:thesis}.

%

\section{Conclusion and Future Work}\label{sec:conclusion}

Our work is an attempt to provide better query support for web
databases having tables with shared attributes using learned attribute
dependencies but missing primary key - foreign key relationship. We
use learned attribute dependencies to make up for the missing PK-FK
information and recover entities spread over multiple tables.  Our
experimental results demonstrate that approach used by \smartint\ is
able to strike a better balance between precision and recall than can
be achieved by relying on single table or employing direct joins.

We are currently exploring a variety of extensions to the \smartint\
system. These include (i) differentiating the importance of the
attributes in tuple expansion (ii) allowing variable width answers, and
assessing the diminishing rewards of additional information using a
discounted reward model and (iii) considering vertical fragmentation
of tables in addition to horizontal fragmentation (which will involve
operating with a set of base tables rather than a single one).

%

\begin{thebibliography}{10}
\providecommand{\url}[1]{#1}
\csname url@samestyle\endcsname
\providecommand{\newblock}{\relax}
\providecommand{\bibinfo}[2]{#2}
\providecommand{\BIBentrySTDinterwordspacing}{\spaceskip=0pt\relax}
\providecommand{\BIBentryALTinterwordstretchfactor}{4}
\providecommand{\BIBentryALTinterwordspacing}{\spaceskip=\fontdimen2\font plus
\BIBentryALTinterwordstretchfactor\fontdimen3\font minus
  \fontdimen4\font\relax}
\providecommand{\BIBforeignlanguage}[2]{{%
\expandafter\ifx\csname l@#1\endcsname\relax
\typeout{** WARNING: IEEEtran.bst: No hyphenation pattern has been}%
\typeout{** loaded for the language `#1'. Using the pattern for}%
\typeout{** the default language instead.}%
\else
\language=\csname l@#1\endcsname
\fi
#2}}
\providecommand{\BIBdecl}{\relax}
\BIBdecl
\smallskip
\smallskip
\bibitem{melnik:simiflood}
\BIBentryALTinterwordspacing
S.~Melnik, H.~Garcia-Molina, and E.~Rahm, ``Similarity flooding: a versatile
  graph matching algorithm and its application to schema matching,''
  in ICDE 2002.

\bibitem{halevy:mlinteg}
A.~Doan, P.~Domingos, and A.~Y. Halevy, ``Learning to match the schemas of data
  sources: A multistrategy approach,'' \emph{Machine Learning}, vol.~50, no.~3,
  pp. 279--301, 2003.

\bibitem{li:semint}
W.-S. Li and C.~Clifton, ``Semint: a system prototype for semantic integration
  in heterogeneous databases,'' in SIGMOD 1995.

\bibitem{larson:theory}
J.~Larson, S.~Navathe, and R.~Elmasri, ``A theory of attributed equivalence in
  databases with application to schema integration,'' \emph{IEEE Tran.
  on Software Engineering}, vol.~15, 1989.

\bibitem{lenz:di}
M.~Lenzerini, ``Data integration: A theoretical perspective,'' in \emph{PODS},
  2002, pp. 233--246.

\bibitem{levy-views}
A.~Y. Halevy, ``Answering queries using views: A survey,'' \emph{The VLDB
  Journal}, vol.~10, no.~4, pp. 270--294, 2001.

\bibitem{emerac-jiis}
\BIBentryALTinterwordspacing
S.~Kambhampati, E.~Lambrecht, U.~Nambiar, Z.~Nie, and G.~Senthil, ``Optimizing
  recursive information gathering plans in emerac,'' \emph{Journal of
  Intelligent Information Systems}, vol.~22, 2004.

\bibitem{lim:entity}
E.-P. Lim, J.~Srivastava, S.~Prabhakar, and J.~Richardson, ``Entity
  identification in database integration,'' apr. 1993, pp. 294 --301.

\bibitem{demichiel:resolving}
L.~DeMichiel, ``Resolving database incompatibility: an approach to performing
  relational operations over mismatched domains,'' \emph{Knowledge and Data
  Engineering, IEEE Transactions on}, vol.~1, no.~4, pp. 485 --493, dec. 1989.

\bibitem{vagelis:discover}
V.~Hristidis and Y.~Papakonstantinou, ``Discover: keyword search in relational
  databases,'' in VLDB 2002.

\bibitem{balmin:objectrank}
A.~Balmin, V.~Hristidis, and Y.~Papakonstantinou, ``Objectrank: authority-based
  keyword search in databases,'' in VLDB 2004.


\bibitem{gaurav:banks}
G.~Bhalotia, A.~Hulgeri, C.~Nakhe, S.~Chakrabarti, S.~Sudarshan, and I.~Bombay,
  ``Keyword searching and browsing in databases using banks,'' ICDE
  2002.
%

\bibitem{mayassam:keyword}
M.~Sayyadian, H.~LeKhac, A.~Doan, and L.~Gravano, ``Efficient keyword search
  across heterogeneous relational databases,''  ICDE 2007.
%

\bibitem{TANEFull}
\BIBentryALTinterwordspacing
Y.~Huhtala, J.~K{\"a}rkk{\"a}inen, P.~Porkka, and H.~Toivonen, ``{TANE}: An
  efficient algorithm for discovering functional and approximate
  dependencies,'' \emph{The Computer Journal}, vol.~42, no.~2,
  1999.

\bibitem{QPIAD}
G.~Wolf, H.~Khatri, B.~Chokshi, J.~Fan, Y.~Chen, and S.~Kambhampati, ``Query
  processing over incomplete autonomous databases,'' in VLDB 2007.

\bibitem{AIMQ}
U.~Nambiar and S.~Kambhampati, ``Answering imprecise queries over autonomous
  web databases,'' in \emph{ICDE}, 2006, p.~45.

\bibitem{CORDS}
I.~F. Ilyas, V.~Markl, P.~Haas, P.~Brown, and A.~Aboulnaga, ``Cords: automatic
  discovery of correlations and soft functional dependencies,'' in
  SIGMOD 2004.

\bibitem{ak:thesis}
A.~Kalavagattu.
Mining approximate functional dependencies as condensed
representations of association rules.   Master's thesis. Arizona State
University. 2008.

\bibitem{smartint-icde}
R.~Gummadi, A.~Khulbe, A.~Kalavagattu, S.~Salvi, and S.~Kambhampati,
  ``Smartint: A system for answering queries over web databases using attribute
  dependencies,'' ICDE  2010 (Demo).

\end{thebibliography}

{\small

}
\end{document}